\documentclass[conference]{IEEEtran}
\usepackage{cite}
\usepackage{amsmath,amssymb,amsfonts}
\usepackage{algorithmic}
\usepackage{graphicx}
\usepackage{textcomp}
\usepackage{color}
\usepackage{xcolor}
\usepackage{lineno}
\usepackage[inline]{enumitem}
\def\BibTeX{{\rm B\kern-.05em{\sc i\kern-.025em b}\kern-.08em
    T\kern-.1667em\lower.7ex\hbox{E}\kern-.125emX}}
\begin{document}

\title{\textsc{serenade}: A Model for Human-in-the-loop Automatic Chord Estimation}

\author{\IEEEauthorblockN{Hendrik Vincent Koops$^\star$$^1$, Gianluca Micchi$^\star$$^1$, Ilaria Manco$^{1,2}$, Elio Quinton$^1$}
\IEEEauthorblockA{
$^1$\textit{Music and Audio Machine Learning Lab, Universal Music Group}\\
$^2$\textit{School of EECS, Queen Mary University of London}\\
London, UK\\
\texttt{\{vincent.koops\},\{gianluca.micchi\},\{ilaria.manco\},\{elio.quinton\}@umusic.com}}}
% \and
% \IEEEauthorblockN{Gianluca Micchi}
% \IEEEauthorblockA{\textit{MAML}\\
% \textit{Universal Music Group}\\
% London, UK\\
% gianluca.micchi@umusic.com}
% \and
% \IEEEauthorblockN{Ilaria Manco}
% \IEEEauthorblockA{\textit{MAML}\\
% \textit{Universal Music Group}\\
% London, UK\\
% ilaria.manco@umusic.com}
% \and
% \IEEEauthorblockN{Elio Quinton}
% \IEEEauthorblockA{\textit{MAML}\\
% \textit{Universal Music Group}\\
% London, UK\\
% elio.quinton@umusic.com}

% For the author list in the Creative Common license, please enter author names.
% Please abbreviate the first names of authors and add 'and' between the second to last and last authors.
\def\authorname{H.V. Koops, G. Micchi, I. Manco and E. Quinton}

% Optional: To use hyperref, uncomment the following.
%\usepackage[bookmarks=false,pdfauthor={\authorname},pdfsubject={\papersubject},hidelinks]{hyperref}
% Mind the bookmarks=false option; bookmarks are incompatible with ismir.sty.

\sloppy % please retain sloppy command for improved formatting

\maketitle
\def\thefootnote{$\star$}\footnotetext{These authors contributed equally.
For the purpose of open access, the author has applied a Creative Commons Attribution (CC
BY) license to any Author Accepted Manuscript version arising.
}\def\thefootnote{\arabic{footnote}}
\begin{abstract}
Computational harmony analysis is important for \textsc{mir} tasks such as
automatic segmentation, corpus analysis and automatic chord label estimation.
However, recent
research into the 
ambiguous nature of musical harmony, causing limited inter-rater agreement,
has made apparent that there is a
glass ceiling for common metrics such as accuracy.
% performance improvements have been hindered by
% a glass ceiling effect caused by several factors, including the potentially
% ambiguous nature of musical harmony and limited inter-rater agreement in
% reference annotations. 
Commonly, these issues are addressed either in the
training data itself by creating majority-rule annotations or during the
training phase by learning soft targets. We propose a novel alternative
approach in which a human and an autoregressive model together co-create a
harmonic annotation for an audio track. After automatically generating harmony predictions, 
a human sparsely annotates parts with low model
confidence and the model then adjusts its predictions following human
guidance. 
We evaluate our
model on a dataset of popular music and
we show that, with this human-in-the-loop approach, harmonic
analysis performance improves over a model-only approach.
The human contribution is amplified by the second, constrained prediction of the model.

% We show that, with this human-in-the-loop approach, harmonic
% analysis performance improves over a model-only approach. We evaluate our
% model on a dataset of popular music and show that the average increase in
% accuracy is larger than the human contribution.

% by introducing X\% of human knowledge at points with low model certainty, we gain X+Y\% on commonly used ACE metrics.

\end{abstract}
\section{Introduction}\label{sec:introduction}

Harmony is a fundamental part of Western music and its analysis has been
studied extensively. Computational harmony analysis is important for a number
of Music Information Retrieval (\textsc{mir}) tasks such as automatic segmentation \cite{structureanalysis}, corpus analysis \cite{burgoyne2011expert,de2011corpus} and
automatic chord estimation (\textsc{ace}) \cite{mcvicar2014automatic,humphrey2015four,mcfee2017structured,chen2019harmony,lopez2021augmentednet}. 
\textsc{ace} is a well-researched topic in
\textsc{mir} that studies how to extract a time-aligned sequence of chords
from a given music signal. Commonly, recent \textsc{ace} systems consist
of some type of audio feature extraction and a deep learning method to
classify segmented audio signal into chord classes.

A well-known problem in creating reference annotations for computational
harmony analysis is disagreement between annotators, which makes it
hard to define a general ``ground truth''. 
Differences in annotations can be the result of personal biases (e.g., an annotator's 
preference for annotating chords in their root position)
or of external constraints (e.g., transcribing a piece for different instrumentation than 
originally intended), or they can stem from ambiguity in the music 
itself \cite{doi:10.1080/09298215.2019.1613436,humphrey2015four,
ni2013understanding,micchi2020not}. 
% In a recording in which notes C, E, and G are combined with a melody
% touching a B, it is up to the annotator whether to include B in the chord
% label (C:maj7) or not (C:maj). Both readings can be considered correct, 
% each one expressing a particular selection of
% the harmonic content of the audio signal. 
These difficulties in creating a ``ground truth'' complicate the 
evaluation of model output and have slowed down the progress in 
computational harmony analysis in the past few years.

In previous research, two main approaches have been proposed to deal with
these challenges. The first approach is to resolve conflict by merging different perspectives.
% For example, reference annotations of popular chord label
% datasets \cite{burgoyne2011expert} are often the result of condensing the
% readings of multiple experts into a single set of labels. The resulting
% files are free from internally conflicting labels, but also reflect a
% specific harmonic point of view. The problem of subjectivity is not limited
% to the collection of reference annotations. 
The effect of training \textsc{ace}
algorithms on such annotations is that their evaluation is bound by a
glass-ceiling effect, above which a model is effectively over-fitted to one
particular harmonic point of view \cite{ni2013understanding,humphrey2015four,DBLP:journals/corr/KoopsHBV17,doi:10.1080/09298215.2019.1613436}.

A second approach to deal with ambiguity and subjectivity is therefore to take
it into account at the model level. For example, 
Koops \emph{et al.} \cite{DBLP:journals/corr/KoopsHBV17} proposed a method to personalize chord labels
for individual annotators by learning a shared (soft) representation of
harmony. This results in a model that performs
beyond a one-size-fits all solution, by providing chord labels that are
specific for a user or task.

When combining the observations that
\begin{enumerate*}[label=\alph*)]
  \item current \textsc{ace} models are performing at the level of inter-annotator agreement 
  \cite{flexer2014inter,humphrey2015four} and
  \item further \textsc{ace} performance improvement depends on the particular user or 
  use case \cite{DBLP:journals/corr/KoopsHBV17}, 
\end{enumerate*}
we argue that one possible third avenue of
\textsc{ace} improvement is a human-in-the-loop approach.
 
It has already been shown that a human-in-the-loop approach can make 
\textsc{mir} systems more useful in real-world scenarios, 
because the perception of a large number of musical dimensions are to some extent
subjective or personal\cite{flexer2014inter}.
For example, Yamomoto \cite{yamamoto2021human} proposes a human-in-the-loop user interface
for beat tracking that interactively adapts to a specific user and target
music. It is argued in \cite{yamamoto2021human} that the interpretation of a beat differs for each
individual, which means that an ideal beat tracking system needs to produce
different outputs from a single input depending on the context. By iterating
interaction between the user and model, the system adapts the internal neural
network model to the user, allowing it to produce more desirable results, while only
requiring the user to change a small portion of the errors that are noticed. 
A human-in-the-loop approach can also help with speeding up labeling 
long audio files, as shown by Kim and Pardo in \cite{kim2018human}. 
For more human-in-the-loop approaches in \textsc{mir} we refer 
to the discussion presented by Yamamoto in \cite{yamamoto2021human}.

% Human-in-the-loop approaches are also getting traction in music creativity.
% The AI Song Contest panel has detailed the process of several groups of researchers
% writing a new piece of music with the help of AI\cite{huang2020ai}. Examples vary: a group reported
% that they let the model create a cappella vocals made of babbling sounds and then
% tried to translate those sounds into actual words. Another group confessed that they took
% as much of the quirky machine output as possible and integrated it into hand-crafted
% songs in a process that constantly refueled their creativity.

% The system adapts the internal model
% to create an annotation that is tuned for a particular user.

% \begin{figure*}
%     \includegraphics[width=\linewidth]{figs/flat_model.png}
%     \caption{
%         Placeholder figure
%     }
%     \label{fig:overview}
% \end{figure*}

\textbf{Contribution.} The contribution of this paper is threefold. 
First, we introduce a novel human-in-the-loop approach to automatic
chord estimation. Our approach is similar to in-painting, in which the
model is responsible for filling in the missing chord labels of sparse human
annotations. More specifically, we simulate a human in the loop by experimenting
with oracles that provide different kinds of harmony information.
Second, we evaluate our model on a dataset
of popular music and we show that with our human-in-the-loop approach we
can improve over a model-only approach, and can produce an output with a 
Return On Investment larger than 1 (cf. the end of Sec.~\ref{sec:humanloop} for a definition of \textsc{roi}).
Third, we extend the model introduced in Micchi \emph{et al.} \cite{micchi2021deep} and
apply it to the audio domain.

 % by introducing X\% of human knowledge at
 % points with low model certainty, we gain X+Y\% on commonly used \textsc{ace}
 % metrics.

\section{Method}
\label{sec:method}

% In this paper, we tackle the task of \textsc{ace} with autoregressive models.
% We first describe the basic architecture we took inspiration from (Section \ref{model}).
% Next, Section \ref{sec:serenade} introduces a new autoregressive model,
% detailing what is novel in our proposal.
% Finally, in Section \ref{sec:humanloop} we discuss our human-in-the-loop approach to \textsc{ace}.

\subsection{An Introduction to \textsc{nade}}
\label{model}
The \emph{Neural Autoregressive Distribution Estimator} (\textsc{nade}) has been proposed 
to model the distribution of random variables that are not independent of each 
other\cite{larochelle2011neural,uria2016neural}. Its core idea is to create 
a hidden layer that gets updated every time a new prediction is made, so that 
subsequent predictions are informed of the previous outcomes.
In simple terms, the \textsc{nade} begins in a state determined uniquely by the biases.
At every step, the logit for the single variable $x[d]$ is predicted based on the value of 
the hidden state, then the variable is sampled, and finally the hidden state is updated 
based on 
% the value of
the sampled output.

The main equations governing the original formulation of \textsc{nade} are the following:
\begin{align*}
  p(x[d]) &= \sigma\big(M_{HV} \sigma(h[d]) + b_v[d])\big)\\
  h[d] &= M_{VH}[:,d-1] x[d-1] + h[d-1]\\
  h[0] &= b_h
\end{align*}
where we use \emph{NumPy}'s notations for arrays\cite{harris2020array},
 $x$ is the array of sampled binary outputs from the visible layer (of size $D$),
 $p(x[d])$ is the probability of sampling 1 at position $d$ for array $x$,
 $h$ is the hidden layer (of size $H$),
 $b_v$ is the bias of the visible layer,
 $b_h$ is the bias of the hidden layer,
 and $M_{HV}$ and $M_{VH}$ are the matrices of weights linking, respectively, 
 hidden layer to visible layer and vice versa (of size $D \times H$ and $H \times D$ 
 respectively).
Teacher forcing can be used to speed up the training: real target outputs, 
instead of the model output, are used as inputs to the next step of the prediction\cite{williams1989learning}.

\textsc{nade} has been sparsely used in \textsc{mir} so far, for example for the task of music 
generation\cite{boulanger2012modeling,huang2019counterpoint} and for the task of harmonic 
analysis\cite{micchi2021deep}. In the latter case, the classification of harmony is done through 
the analysis of fixed-length frames of music through a convolutional recurrent neural network 
(\textsc{crnn}), followed by a \textsc{nade} that replaces the more standard shallow classifiers.
The connection between the \textsc{crnn} and the \textsc{nade} is done, 
following Boulanger \emph{et al.}~\cite{boulanger2012modeling}, by using the output of the recurrent network to determine 
the biases of the \textsc{nade}. At each timestep, the label is divided into 6 multiclass 
sub-labels: key, tonicisation, \emph{droot} (distance between key root and chord root), quality, inversion, and root. The basic architecture 
of \textsc{nade} is therefore extended to predict a multiclass variable instead of a single 
binary output. 
For example, and as a consequence of the architecture, the choice of the quality of the chord 
at timestep $t$ was informed by the sampled key, tonicisation, and \emph{droot}, therefore avoiding 
incoherent outputs.

\begin{figure}
    \includegraphics[width=\linewidth]{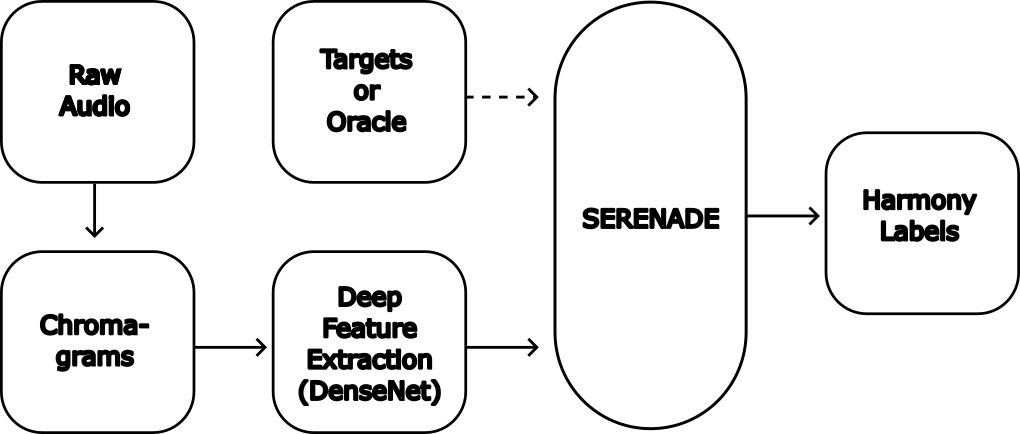}
    \caption{
        A schematic look at the model architecture.
        The targets/oracle connection can be absent when the model is performing a 
        standard inference, hence the dashed lines.
    }
    \label{fig:model}
\end{figure}

\subsection{\textsc{serenade}: A Separable \textsc{nade} for Harmony Analysis}
\label{sec:serenade}

We modify and extend the \textsc{nade} architecture presented in Micchi \emph{et al.} \cite{micchi2021deep} and introduce \textsc{serenade}.
The main novelties are:

 \textbf{Audio.} 
 We apply the \textsc{nade} to audio content instead of symbolic content.
 Our audio files are segmented into 60-second excerpts. Chromagrams for both the bass and global 
 frequency spectrum of the music are extracted using \emph{NNLS Chroma}, following Burgoyne \emph{et al.}~\cite{burgoyne2011expert}.
 The chromagrams are then used as inputs to the \textsc{crnn} part of the network, 
 implemented with a 1D DenseNet architecture\cite{huang2017densely}, acting as a deep feature extractor.
 
\begin{figure}
    \includegraphics[width=\linewidth]{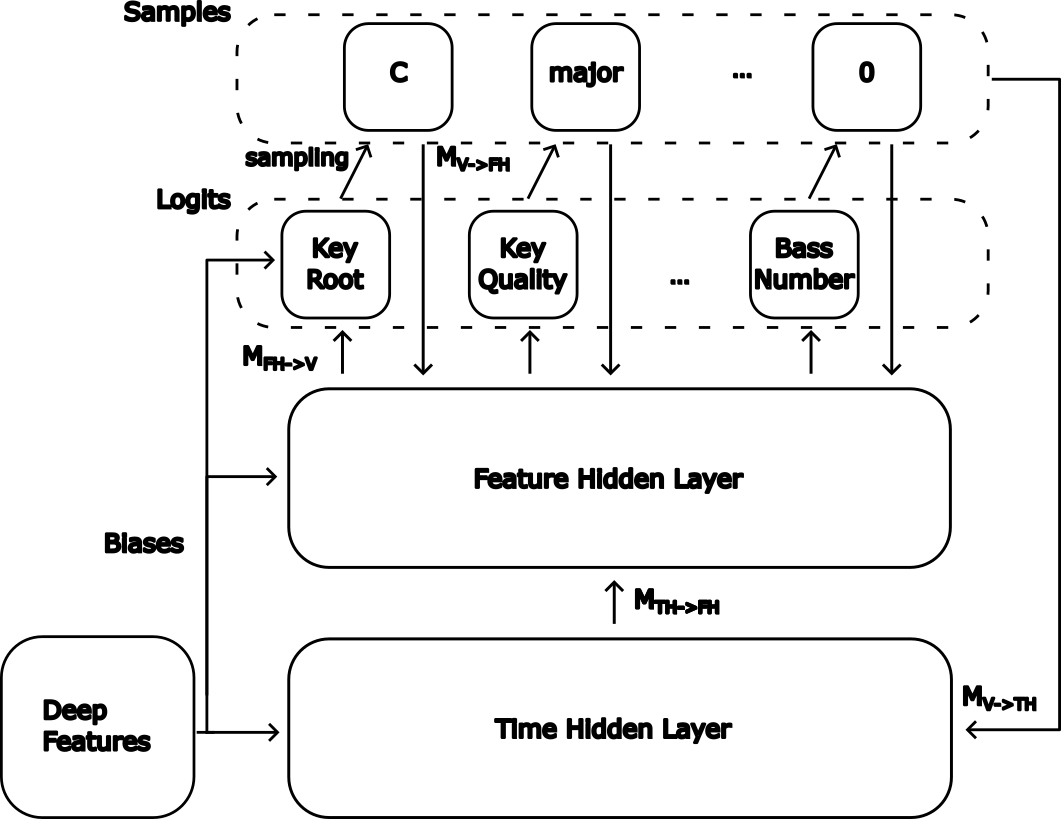}
    \caption{
        Our proposed \textsc{serenade} model in the ``infer'' configuration.
        When training, logits are not sampled and instead the targets are used to take 
        advantage of the teacher forcing paradigm.
        When using an oracle, instead, either ground truth or samples are used 
        based on the value of an oracle mask: if the mask is active, take the ground truth; 
        if inactive, sample from the logits. In all three configurations, the biases change 
        at every time step.
        The final version of the model is bidirectional, which means that two identical 
        \textsc{nade} are used, one going from the past to the future and the other vice versa.
        Their logits are then averaged before sampling.
    }
    \label{fig:nade}
\end{figure}

 \textbf{Depth.} 
 We extend the \textsc{nade} mechanism to the temporal dimension.
 To achieve this, we replace the \textsc{rnn} with an additional \textsc{nade} hidden layer, responsible for modelling the harmony autoregressively along the time dimension.
 The output of the deep feature extractor determines the biases of all three \textsc{nade} layers 
 (two hidden and one visible);
 the first hidden layer (or \emph{time} hidden layer) is used to determine the value of the 
 second hidden layer (or \emph{feature} hidden layer) and gets updated once per time step;
 the feature hidden layer is used exactly as before and is updated at every sampling of a sub-label.
 We refer to this approach as \textit{separable}, since time and feature dimensions are modelled separately (see Fig.~\ref{fig:nade}).
 An alternative to this would be to flatten both dimensions and model them with a single hidden layer.
 Through preliminary experiments we verify that the former performs better and therefore only consider the separable variant in the rest of the paper.
 The model is made bidirectional by defining two parallel \textsc{nade}s, one left-to-right 
 and one right-to-left, the logits of which are then averaged before sampling at inference time.
 
 \textbf{Directed Root (\emph{droot}) Label.} 
We simplify the task of determining the \emph{droot} label by adding a constant value of $3$ to the logits of the calculated \emph{droot} 
in order to direct the model towards the right choice (see Section~\ref{sec:preprocessing} for further details).
The value of $3$ was chosen empirically by noticing that typical logits for a confident prediction were ranging 
between $-5$ and $+5$: $3$ thus represents a strong nudge in the right direction of the same order of magnitude 
of all other effects in the model.
% \item \textbf{Pop music.} We use a different set of harmonic sub-labels, which is more appropriate to the popular music context we apply the model to.
% key root, key quality, chord root, \emph{droot}, quality, and bass.

The proposed architecture is shown in Figures~\ref{fig:model} and~\ref{fig:nade}.
The \textsc{serenade} has 175k weights and can be trained on a single GPU in about one hour.

\subsection{Human-in-the-loop Simulation}
\label{sec:humanloop}
We simulate a human in the loop as an oracle that provides a sparse annotation for certain sub-labels 
at given time steps (for brevity, we sometimes extend the name oracle to the sparse annotation itself).
The oracle can make independent choices of time steps for different sub-labels.
For example, one can provide the chord root at frames 48 and 102 and the quality at frame 55.
In all cases, whenever the oracle is active, it is used both to overwrite the output sampled by the model 
and also to update the hidden state of \textsc{serenade} by means of teacher forcing.
To help us in the evaluation of the oracle experiments we define two new quantities.

\textbf{Oracle cost.}
The \emph{oracle cost} is a measure of how many annotations we have provided an oracle for.
It is expressed as a percentage with respect to the total number of annotations in the audio extract under analysis.
For example, providing the oracle for the key root at all time steps has a cost of 16.67\% because the key root is one of 6 sub-labels.
If we were to provide it only half the times the cost would drop to 8.33\%.

\textbf{Oracle \textsc{roi}.}
The second quantity is the \emph{oracle \textsc{roi}}, or Return On Investment.
The \textsc{roi} is defined as the ratio between the change in accuracy due to the oracle and the oracle cost:
\begin{equation*}
    \textsc{roi} = \frac{\text{accuracy}_\text{oracle} - \text{accuracy}_\text{original}}{\text{oracle cost}}
\end{equation*}

In a model without autoregression, an oracle always has an \textsc{roi} between 0 and 1.
It is 0 if all the labels provided were already correctly predicted by the model, 
1 in the opposite (ideal) case where the oracle was used only on wrong predictions,
and in between these extremes for all other situations.
This is no longer true for \textsc{serenade}.
Due to its autoregressive nature, a correction at one specific sub-label can be propagated to 
different sub-labels at the same time step and also to other time steps.
A possibility for \textsc{roi} larger than 1 is therefore unlocked.
This means that the model can make human corrections more effective and, for example, considerably 
speed up the annotation of new datasets.
It is maybe not trivial to notice that, in theory, also negative \textsc{roi}s can happen, when 
the oracle is actually detrimental to the global quality of the output. 
However, these cases are extremely unlikely and never occurred.
% Section~\ref{sec:oracleresults} provides the results of our experiments with oracles.

\begin{figure}
    \includegraphics[width=\linewidth]{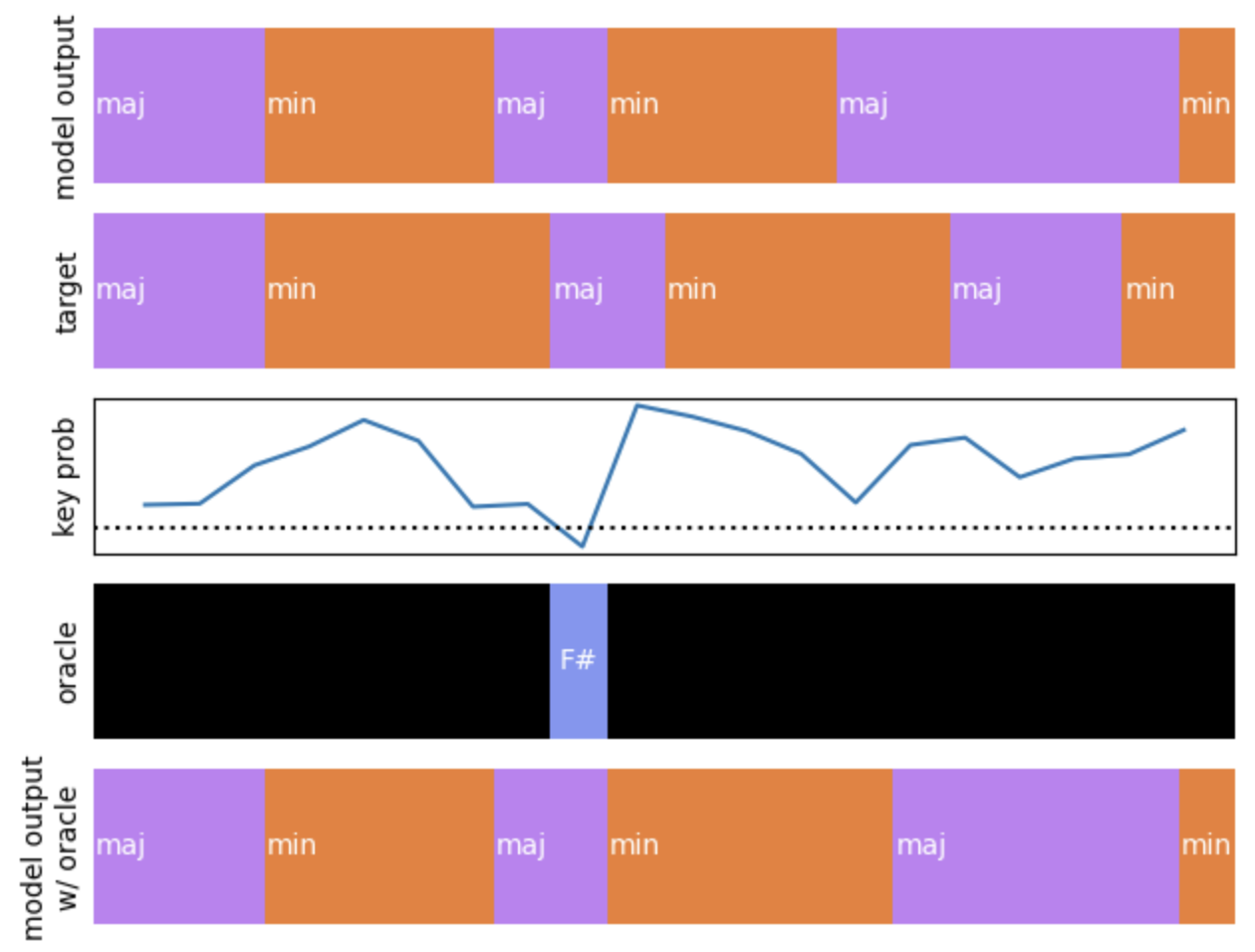}
    \caption{
        Our simulated human-in-the-loop approach.
        We analyse chord quality and key root at the same time to show the effects across sub-labels,
        % that one sub-label has on the others,
        even at different time steps.
        We see that the confidence of the model in the key root is below the threshold 0.35.
        Providing an oracle when going below that threshold has a positive impact on the quality of the chord, which
        improves total accuracy. This is an example of \textsc{roi} $>$ 1.
        See Section~\ref{sec:oracleresults} for more details.
    }
    \label{fig:oracle}
\end{figure}

 \begin{table*}[t]
 % \resizebox{\textwidth}{!}{% use resizebox with textwidth
 \centering
  \caption{Harmonic class labels used for training our models. 
   The \emph{droot} values represent the semitone index relative to the key 
   (for example the chord \texttt{C:maj} in the key of \texttt{C} \texttt{major} will have a \emph{droot} of \texttt{0}, 
   and \texttt{G:maj} in the same key will be \texttt{7}.)
   The values \texttt{-1} and \texttt{N}
   represent a musical passage without harmonic annotations (silence, for
   example).
   Pitches not mentioned in this table are mapped to their enharmonic equivalents.}
  \label{tab:data}
\begin{tabular}{l|l}
Type & Possible Values  \\
\hline
Key root & \texttt{\{C, C\#, D, D\#, E, F, F\#, G, G\#, A, A\#, B, N\}} \\

Key quality & \texttt{\{major, minor, N\}}  \\
Chord root & \texttt{\{C, C\#, D, D\#, E, F, F\#, G, G\#, A, A\#, B, N\}}  \\
\emph{Droot} & \texttt{\{0, 1, 2, 3, 4, 5, 6, 7, 8, 9, 10, 11, -1\} }  \\
Chord quality &  \texttt{\{maj, min, dim, aug, 7, maj7, min7, 5, 1, sus4, N\}}  \\
Bass number & \texttt{\{0, 1, 2, 3, 4, 5, 6, 7, 8, 9, 10, 11, -1\} }  
\end{tabular}
% }
\end{table*}

\section{Dataset preparation}
\label{datasets}

We use two datasets as ground-truth annotations 
(with all the caveats discussed in the introduction) for our
experiments: the \emph{Isophonics} annotations for \emph{The Beatles} and a subset of 
the \emph{Billboard} dataset.
% for which we have the rights.
Both datasets contain time-aligned key and chord label reference annotations.
In the following sections, we briefly detail the contents of these datasets,
followed by a description of our data pre-processing approach. 
We follow the commonly used chord label notation convention
introduced by Harte et al. in Harte \emph{et al.} \cite{harte2005symbolic}.

\textbf{Billboard.} This dataset was introduced by Burgoyne et al.
 in \cite{burgoyne2011expert} and contains chord label annotations for songs
 sampled from the \emph{Billboard ‘Hot 100’} music charts. The annotations
 are the result of the consensus of expert annotations, and are a staple
 dataset for computational harmony analysis and automatic chord estimation of
 popular music. This dataset contains reference annotations for 890 songs. In
 this paper we use a subset of 267 songs that are accessible to us without
 copyright restrictions.
 % Originally, this dataset contains reference annotations for 890 songs.
 % To prevent training our models on data we do not have the rights for, we
 % subsample this dataset. To achieve this, we look up all the ISRCs of the
 % tracks in the \emph{Billboard} dataset, and find its respective rights' owner. If
 % we are the owner, we keep the track, otherwise we discard it. This filtering
 % process leaves us with 267 songs.

\textbf{Isophonics -- The Beatles.} This dataset comprises 178 tracks from The Beatles 
catalog with annotations relating to
chords, keys, structural segmentations, and beats/bars. These annotations were
 collected by the Centre for Digital Music (C4DM) of Queen Mary University of
 London\cite{mauch2009omras2}. For the experiments in this paper, we
 leverage the chords and key annotations.

\subsection{Data Pre-Processing}
\label{sec:preprocessing}
By combining the \emph{Billboard} and \emph{Beatles} datasets we obtain a training 
dataset with 445 songs. We resample the audio at 22,050 Hz and create two files 
for each song: chromagram features and harmony labels.

\textbf{Audio Features: Chroma.}
We extract \emph{NNLS Chroma} features, introduced by Mauch et al. in \cite{mauch2010approximate}.
Chroma features capture the pitch-class content of harmony in terms of the 12 pitch classes folded into a single octave.
They are commonly used for automatic chord estimation since they provide a
 good proxy for capturing harmonic information \cite{gomez2006tonal}.
 In addition to the standard set of 12 chroma features analysing the entire spectrum of frequencies,
 we also produce another set of 12 containing only the frequencies from the lower octaves.
 The distance in milliseconds between two successive frames in the chromagram is roughly 46 ms.
 The specific implementation closely follows the \emph{Billboard} dataset \cite[Appx. B]{burgoyne2012stochastic} for compatibility reasons.
% This is because lower octaves give us information on the bass of the chord, which has a special role in harmony.
 % We extract these 24 features with the NNLS chroma VAMP plugin for sonic annotator \cite{chris2010a}.
 % We use the default plugin settings with
 % the exception for the recommended 1\% rolloff for pop music.
 % The choice of the type of features and its implementation was done to keep compatibility
 % with the \emph{Billboard} dataset \cite[Appx. B]{burgoyne2012stochastic}.
 
\textbf{Audio Features: Pitch Profiles.}
 Chroma features are extremely localised in time: each frame only describes a few tens of ms of audio.
 Some harmonic concepts, such as the key, are usually defined over much larger time scales.
 To better bridge this gap, we augment the data with global pitch profiles computed from the average of each of the 12 main chroma features over the entire excerpt of audio used for analysis (60 seconds).
 We then take the inner product of these additional 12 values with each of the 24 pitch profiles proposed by Temperley for key detection\cite{temperley1999s}.
The result is then concatenated to the extracted chroma features at each time step. 
By including it, we can simultaneously provide the model with local harmony (chroma) and global harmony (pitch profiles) information. Previous research in 
\textsc{ace} has shown that providing other musical parameters can improve chord 
estimation results \cite{mauch2009using}.

% which we hypothesize will benefit both chord label and key estimation.
% This has vastly improved the results in key detection, which jumped from 48\% to over 80\% accuracy on the test dataset.

% \begin{figure*}
%   \centering
%   % \includegraphics[width=\textwidth,trim={5cm 5cm 8cm 1cm},clip]{figs/data.pdf}
%   \includegraphics[width=\textwidth,trim={2.7cm 11cm 7.5cm 2.5cm},clip]{figs/data_smaller.pdf}
%   \caption{Example data annotations used in our experiments. On the left, the
%    figure shows the bass and full spectrum chroma used in our experiments.
%    Next to it, the figure shows the key and chord annotations for the
%    corresponding chromagram frames. This example visualizes the first couple
%    of seconds of \textit{Let it Be} by the Beatles.
%    VINCENT: CHANGE FIGURE INDEXES!
%    }
%   \label{fig:data}
% \end{figure*}

\textbf{Labels: Harmony Annotations.} For each track, we create a new
 reference annotation that is aligned with the extracted chromagrams. In
 these annotations, we represent the harmony with six sub-labels: 
 key root (e.g. \texttt{C}), 
 key quality (e.g. \texttt{major}),
 chord root (e.g. \texttt{A}),
 \emph{droot} (or \emph{directed distance to root}, which is the name that we give to the distance in semitones between the chord root and key root),
 chord quality (e.g. \texttt{minor}),
 and bass number (the distance in semitones between the bass note and the root of the chord).
 Table \ref{tab:data} provides an overview of all harmonic class labels used in our experiments.

\textbf{Labels: \emph{Droot}.}
The \emph{droot} is uniquely determined by the chord root and key root.
Hence, it is redundant and could be removed from the set of predictions.
However, \emph{droot} and quality are related:
for example, the base triad of a chord written on the dominant is (almost) always major, 
while one built on the subdominant is usually major in major keys and minor in minor keys.
Therefore jointly predicting both is likely to provide an overall net benefit to the model 
performance.
A small positive effect has been \emph{The Beatles} by our experiments in Section~\ref{sec:oracleresults}.

\textbf{Labels: Key Quality.}
 The \emph{Billboard} dataset does not include information on the key quality.
 To estimate it, we look at the predominant quality of tonic chords:
 if they are mostly minor we assume it's a minor key;
 vice versa, we assume it's major.
 This excludes other modes such as dorian, aeolian, mixolydian, but we believe that
 the additional information provided by this simple inferral scheme
 outvalues the noise introduced in the dataset.
 For the \emph{Beatles} tracks, we use the ground-truth annotations provided by the dataset, 
 but exclude tracks were the key is modal (i.e. other than \texttt{major} or \texttt{minor}).

\textbf{Labels: Chord Quality.}
To prevent an explosion of possible
 chord label combinations and reduce the effect of infrequently occurring
 chord types, we reduce our chord qualities to one of 11 classes:
 \texttt{maj} (60.8\% of annotations),
 \texttt{min} (15.1\%),
 \texttt{dim} (0.6\%),
 \texttt{aug} (0.3\%),
 \texttt{7} (6.7\%),
 \texttt{maj7} (2.2\%),
 \texttt{min7} (6.2\%),
 \texttt{5} (1.8\%),
 \texttt{1} (1.4\%),
 \texttt{sus4} (2.8\%),
 and \texttt{N} (1.9\%).
To do so, chord qualities \texttt{maj6, 9, 11, 13, maj9, maj13}, and \texttt{sus2}, 
totalling 3.1\% of the annotations, are mapped to \texttt{maj} (57.7\%).
Qualities \texttt{min6, minmaj7, min9, min11}, and \texttt{min13}, totalling 0.9\% of the 
annotations, are mapped to \texttt{min} (14.2\%).
Finally, we combine \texttt{dim, dim7}, and \texttt{hdim7}.

The chord quality mapping was made with music theory concerns in mind, trying to 
aggregate chords that are similar to each other either in terms of pitch-class content 
or of compositional usage.
We recognize that no chord reduction is optimal in all settings.
One could even argue that all chord reductions are wrong to some degree.
However, mapping chord labels to a smaller number of classes
is a commonly used approach in automatic chord estimation to remedy the
combinatorial explotion of chord labels, and to facilitate evaluation and
model training \cite{raffel2014mir_eval}.

In the following sections, we discuss our experiments and results.
First we evaluate our model in a way that is common in \textsc{ace} research in Section 
\ref{sec:baseline}. In these experiments, we use common \textsc{ace} metrics and compare 
our results with the current state of the art.
% After these results, we introduce in Section \ref{sec:oracleresults} our human-in-the-loop 
% evaluation method. We explain how we measure the return on investment of our oracle intervention  
% and present results for various intervention approaches.
Then, in Section~\ref{sec:oracleresults}, we focus on the experiments we have run to test
our human-in-the-loop approach.

\section{Results}
\subsection{Baseline Results on Automatic Chord Estimation}
\label{sec:baseline}
% We use harmonic-analysis-training-2023-04-09-06-57-48-673 for our results
% Before detailing the experiments in annotations with a human in the loop we make a recap of the
% quality of this model when used in the automatic fashion.

\begin{figure}    \includegraphics[width=\linewidth]{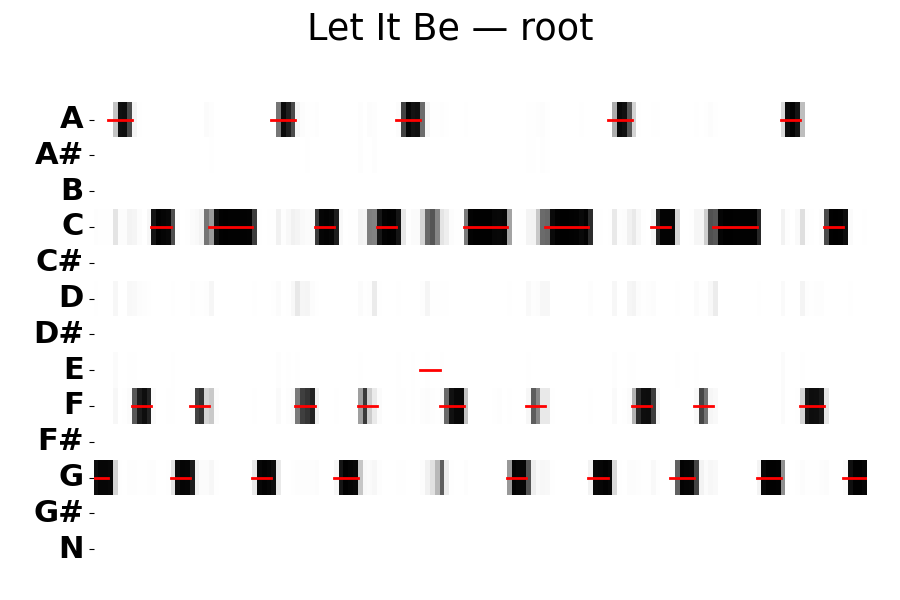}
    \caption{The chord root prediction of \textsc{serenade} (heatmap, in greyscale) follows the ground truth (horizontal red lines) quite closely.}
    \label{fig:root}
\end{figure}

Fig.~\ref{fig:root} shows the chord root prediction for 60 seconds of 
\emph{``Let It Be''}, by The Beatles.
Overall the predictions are accurate, except that most of them arrive one or two frames later than the 
ground truth. This could either be due to a limitation of our model or to a small alignment 
issue in the ground truth.

The performance of our model compared to top performing entries in the \textsc{mirex} evaluation\cite{mirex_website} is shown in Table~\ref{table:results}.
All metrics are computed using the python package \texttt{mir\_eval}\cite{raffel2014mir_eval}.
As a caveat, the \textsc{mirex} evaluation dataset is not publicly available so we are evaluating 
on a different subset of 61 songs coming from the \emph{Billboard} v2.0.
Our proposed model is the best performing in terms of root detection.
The \textsc{majmin} score is competitive and, adding a key root oracle (see Section~\ref{sec:oracleresults}) it outperforms the baselines.
The \textsc{sevenths} score, however, shows room for improvement.
We hypothesise the difference with JLCX2 be due to the training dataset size or choice encoding 
for qualities, which treats every common extension to the triads as a separate binary unit.
Such a scheme seems to be a good candidate for being applied also to our \textsc{serenade} model 
and we are considering it for further experiments.

\begin{table}
    \caption{
        Comparison between our model and others that were evaluated as part of the \textsc{mirex} competition.
        [wko] = with key oracle, as described in Experiment~1.
        *: the results for our model \textsc{serenade} are obtained on a different subset of the \emph{Billboard} 
        dataset as the one used for \textsc{mirex} competitions is not publicly available.
        \textsc{serenade} has been trained on a smaller dataset than others, due to copyright reasons.
    }
    \begin{center}
    \begin{tabular}{l|c|c|c}
        Model & \textsc{root} & \textsc{majmin} & \textsc{sevenths} \\
        \hline\hline
        % JR2 & 62.95 & 57.95 & 43.71 \\
        % HL2 & 65.62 & 59.10 & 48.41 \\
        % PP3 & 70.61 & 67.82 & 50.31 \\
        % NG1 & 70.98 & 66.96 & 48.54 \\
        % CF2 & 71.16 & 67.28 & 48.99 \\
        % CM2 & 71.23 & 67.36 & 49.07 \\
        % DK1 & 72.06 & 68.69 & 54.54 \\
        CLSYJ1 & 73.69 & 69.47 & 57.41 \\
        NMSD2 & 74.72 & 71.40 & 59.30 \\
        WL1 & 75.22 & 72.53 & 57.87 \\
        JLW2 & 75.40 & 72.77 & 57.49 \\
        CB4 & 76.29 & 72.01 & 59.15 \\
        KO2 & 76.52 & 73.82 & 60.21 \\
        SG1 & 77.92 & 72.41 & 58.26 \\
        JLCX2 & 79.38 & 77.92 & \textbf{64.22} \\
        KBK2 & 80.60 & \textbf{78.27} & 55.81 \\
        \textsc{serenade}* & \textbf{80.89} & 77.57 & 57.82 \\
        \textsc{serenade}* \textit{[wko]} & \textit{80.78} & \textbf{\textit{79.20}} & \textit{59.13} \\
    \end{tabular}
    \end{center}
    \label{table:results}
\end{table}

\subsection{Testing the Human-in-the-Loop Approach Using Oracles}
\label{sec:oracleresults}
To test the capabilities of the model to adapt its predictions to human input we develop a few experiments involving oracles.
We run all the experiments on a randomly selected test set made of 30 recording from the \emph{Beatles} 
dataset and 61 from the \emph{Billboard} dataset which was set apart before training our models.
When comparing results to \textsc{mirex}, only the \emph{Billboard} data is used.

\subsubsection{Oracles on Key Root and Key Quality}
Key root and key quality are usually relatively inexpensive oracles to provide in terms of human 
resources as they tend to be constant for long stretches of time, with a majority of pop songs 
having only one key for the entire piece.

When providing these two oracles, we see that there is a global improvement on the standard 
\texttt{mir\_eval} metrics of majmin and sevenths, as shown in Table~\ref{table:results}.
Such a result is qualitatively impossible with standard shallow classifiers instead of 
\textsc{nade} due to the lack of direct transfer of information between the sub-labels.
This result was already reported in~\cite{micchi2021deep}.

\subsubsection{Oracle on \emph{Droot}}
Our hypothesis is that the \emph{droot} can help in determining the 
quality of the chord and we test this by adding the \emph{droot} oracle.
We observe a small performance improvement: the quality prediction accuracy over the test dataset is 66.9\% in the absence of 
any oracles and 67.2\% with the \emph{droot} oracle. 

A detailed oracle on the \emph{droot} is very expensive to make and amounts to basically producing a 
full harmonic annotation by hand, making this avenue impractical to follow for speeding up chord 
annotations. However, it validates our choice of using expert knowledge to define the \emph{droot}
in order to improve the \textsc{ace} quality.

\subsubsection{Oracles Where the Model is Wrong}
Given the knowledge of the ground truth at all times, one can devise an oracle that is active only 
on a subset of the wrong predictions.
For example, when providing the correct label for 10\% of the wrong predictions on root, 
quality, and bass number, we obtain an average \textsc{roi} of 1.18 over all tracks in the dataset,
while the \textsc{roi} for each individual audio excerpt ranges from 1 to 2.5.

This \textsc{roi} is not particularly large for practical purposes but the mere fact that it is 
larger than 1 is a proof of concept that a model optimised to do so can be devised and used for 
faster data labeling.
Also, it is important to notice that this scheme is not unique to the task of harmonic analysis 
and could be generalised to other tasks, within and outside \textsc{mir}.

\subsubsection{Oracles on Predictions with Low Model Confidence}
In some real-world scenarios we may not have a priori knowledge of which predictions are wrong.
For the oracle to be practically useful in such cases, we require a scheme to automatically determine when the annotator's intervention is most likely to be required.

\begin{figure}
    \includegraphics[width=\linewidth]{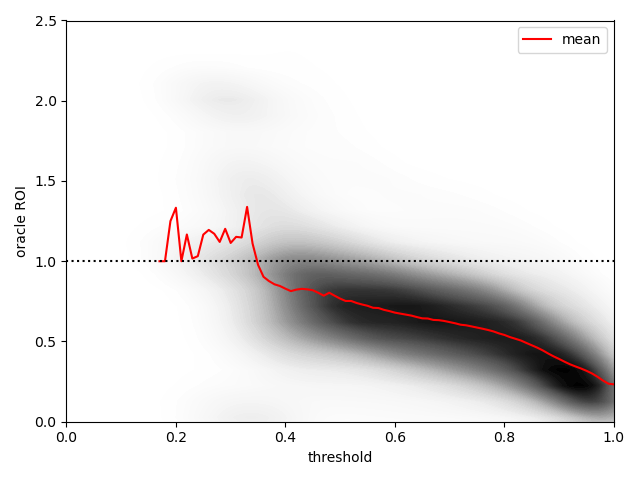}
    \caption{
        The distribution of the \textsc{roi}s for the oracle.
        The red line represents the average of the distribution as a function of the confidence threshold.
        The black shadow is the Kernel Density Estimate (KDE) of the distribution, calculated with the python package Seaborn.
    }
    \label{fig:ROIs}
\end{figure}

We propose to simulate this scenario by providing an oracle whenever the model outputs a prediction with a 
confidence smaller than a threshold $t$.
We then study the distribution over the test dataset of the \textsc{roi}s we obtain as a function  of $t$ 
and plot the results as a heatmap in Fig.~\ref{fig:ROIs}.
We can see that there is a \textsc{roi} $>$ 1 for thresholds roughly between 0.2 and 0.35.
Below 0.2, the oracle is not used because the model never outputs such low confidence values.
Above 0.35, the \textsc{ROI} $<$ 1 reveals that the oracle just confirms a prediction that is already correct.

The assumption that the outputs with low probability are the ones that are most likely 
to be wrong is verified: a threshold up to 0.35 always provides \textsc{roi} larger than 1.

\section{Discussion and Conclusions}
In this work, we have introduced \textsc{serenade}, a first approach to human-in-the-loop automatic 
chord estimation.
We showed that by leveraging the autoregressive nature of our model, we can in-paint partial 
annotations from an oracle to obtain higher-quality predictions overall.
We defined two new quantities to evaluate the results: an oracle cost and an oracle \textsc{roi}.
Our experiments show an average \textsc{roi} of 1.18, meaning that for every single oracle 
annotation provided, 1.18 labels are corrected.
This positive feedback, if harnessed correctly, could eventually lead to a significant speedup in 
the generation of new annotated datasets. For example, in an annotation task a user would only have to provide a corrected key for a piece to improve \textsc{ace} results overall.

We identified two main avenues for future work.
Firstly, assessing  the long-term coherence of the model.
For example, if an oracle provides a correct annotation for the first appearance of the chorus, 
will the model be able to apply that correction to all choruses?
Recent work in generative models showed promising long-term coherence modelling improvenent by employing hierarchical architectures and representations \cite{9533461, agostinelli2023musiclm}, which may also be beneficial to \textsc{serenade}.

Secondly, improving the human-in-the-loop formalism.
In its current form, the oracle cost may not be a fair representation of the actual effort provided by a human annotator. In particular, it treats frames independently and all harmonic elements (such as key root, key 
quality, chord root, etc.) equally. As a result the cost of annotating 10 frames is 10 times greater than annotating one frame, whereas it would likely be "cheap" for a human to annotate segments of many contiguous frames. 
We expect that with a refined oracle cost metric the benefit of such a human-in-the-loop approach may become even more apparent. 

%Currently, the oracle cost is sometimes too simple a measure of the effort that a human annotator would do. In its current formulation, it treats all harmonic elements (such as key root, key quality, chord root, etc.) equally. This approach does not accurately reflects the human effort needed to annotate different harmonic elements. For example, it providing the key for a song, even with modulations, is not as expensive as providing all the chord roots. 
%However, currently, it boils down to the same \emph{oracle cost}.
%We plan to explore a cost that looks at how many different labels one writes down, rather than at how many frames each label covers.

\section{Acknowledgements}
This work was supported by UK Research and Innovation [grant number EP/S022694/1] and Universal Music Group.

\bibliographystyle{IEEEtran}
\bibliography{IEEEabrv,serenade}

\end{document}